\documentclass[screen,nonacm]{acmart}
\AtBeginDocument{%
  }
\usepackage{multirow} 
\usepackage{xcolor,soul}
\usepackage{subcaption}

\begin{document}
\title{Reassessing the Feasibility of PPG-Based Non-Invasive Blood Glucose Level Estimation}


\author{Supraja Ramesh}
\email{supraja.ramesh@kit.edu}
\affiliation{%
  \institution{Karlsruhe Institute of Technology}
  \city{Karlsruhe}
  \country{Germany}
}
\orcid{0009-0007-3736-144X}

\author{Markus Neufeld}
\email{markus.neufeld@student.kit.edu}
\affiliation{%
  \institution{Karlsruhe Institute of Technology}
  \city{Karlsruhe}
  \country{Germany}
}
\orcid{0009-0002-7876-3414}

\author{Michael K{\"u}ttner}
\email{michael.kuettner@kit.edu}
\affiliation{%
  \institution{Karlsruhe Institute of Technology}
  \city{Karlsruhe}
  \country{Germany}
}
\orcid{0009-0000-9021-0359}

\author{Tobias R{\"o}ddiger}
\email{tobias.roeddiger@kit.edu}
\affiliation{%
  \institution{Karlsruhe Institute of Technology}
  \city{Karlsruhe}
  \country{Germany}
}
\orcid{0000-0002-4718-9280}

\author{Michael Beigl}
\email{michael.beigl@kit.edu}
\affiliation{%
  \institution{Karlsruhe Institute of Technology}
  \city{Karlsruhe}
  \country{Germany}
}
\orcid{0000-0001-5009-2327}

\renewcommand{\shortauthors}{Ramesh et al.}
\newcommand{\Vitaldb}{\text{VitalDB}}
\newcommand{\must}{\text{MUST}}
\newcommand{\bglEst}{BGL estimation}

\newcommand{\xgbVitalDB}{\text{ }}
\newcommand{\tinyCNNVitalDB }{\text{MUST}}
\newcommand{\plsVitalDB}{BGL estimation}
\newcommand{\resnetVitalDB}{BGL estimation}
\newcommand{\xgbMFCCVitalDB}{BGL estimation}
\newcommand{\hlBlue}[1]{{\sethlcolor{blue!20}\hl{#1}}}

\newcommand{\tr}[1]{{\sethlcolor{blue!20}\hlBlue{[Tobi: #1]}}}
\newcommand{\sr}[1]{{\sethlcolor{orange!20}\hlBlue{[Supraja: #1]}}}
\newcommand{\mk}[1]{{\sethlcolor{yellow!20}\hlBlue{[Michael K: #1]}}}

\begin{abstract}
Non-invasive blood glucose level (BGL) estimation from photoplethysmography (PPG) holds great promise for wearable health monitoring, but results across studies are hard to compare due to inconsistent datasets, data leakage, and non-standardized evaluation metrics. We present the first reproducible, extensible evaluation pipeline and use it to reassess five representative PPG-based BGL methods on published datasets under three increasingly strict data-split protocols: random window-level, participant-aware, and leave-some-participants-out (LSPO). Models appeared competitive under random splitting but collapsed under participant-aware and LSPO evaluation, with nearly all yielding near-zero or negative R² values comparable to a mean-prediction baseline. Critically, across every model and split, over 90\% of predictions fell within clinically acceptable zones (Clarke Error Grid A+B), including the baseline. This reveals a fundamental disconnect: clinical zone metrics systematically conceal model failure in this domain. Our findings demonstrate that random train–test splits substantially overestimate the generalization of PPG-based BGL models due to sample-level data leakage, and that robust ML evaluation must precede clinical validation to meaningfully assess real-world utility.
\end{abstract}




\maketitle

\section{Introduction}

Diabetes affects millions of people worldwide each year, posing a major challenge to global public health systems \cite{unknown2020quality}. The consequences of diabetes transcend the individual level, manifesting across familial structures and exerting a wide-ranging impact on society at large \cite{WHOdiabetes}. Maintaining blood glucose levels within the normal range is critical; chronic dysregulation can lead to severe complications affecting vital organs and bodily functions \cite{krause2023type}. While lifestyle interventions such as diet regulation and physical activity play key roles in diabetes management \cite{muntis2023high, d2023moderate}, regular blood glucose monitoring has proven helpful in maintaining it \cite{parsons2019effect}.

Traditional self-monitoring techniques are predominantly invasive, requiring finger-prick blood samples. The fear of needles, pain, and the cost of the device are among the main barriers to self-monitoring \cite{ong2014barriers}. While continuous glucose monitoring (CGM) systems represent a step forward, their sensors must be inserted under the skin and typically need replacement every 6–14 days, which can cause discomfort such as wear-related erythema, itchiness, and induration, leading to discontinuation of use \cite{asarani2020cutaneous}. These drawbacks underscore the need for accurate, non-invasive glucose monitoring solutions.

Despite a decade of active research, no non-invasive method has reached commercial viability, largely due to unresolved challenges in clinical reliability, cross-individual consistency, and usability \cite{shang2022products, tang2020non}. Among emerging approaches, photoplethysmography (PPG) has shown promise for estimating blood glucose levels (BGL) using machine learning (ML) techniques \cite{jiang2025ppg}. PPG’s prevalence in wearable technology makes it a strong candidate for continuous, non-invasive glucose monitoring, especially in edge-computing scenarios where computational resources are limited \cite{zeynali2025non}. Even though several studies have demonstrated encouraging results \cite{habbu2019estimation, zeynali2025non, castro2023three}, the lack of uniform evaluation on published datasets and standardized evaluation metrics makes it difficult to compare models fairly and assess their practical feasibility.

In this work, we make two main contributions to advance PPG-based \bglEst{}: (i) We introduce a reproducible, extensible evaluation pipeline \cite{pipeline} that integrates two published datasets (\Vitaldb{} \& \must{}), enabling fair cross-study comparisons under common datasets. (ii) We show, through a systematic evaluation of five representative methods under three increasingly strict data-split protocols, that commonly used random-split evaluation substantially overestimates model generalization and that widely used clinical metrics can mask model failure in this application. Our findings provide direct implications for how future PPG-based \bglEst{} work should be evaluated to have real-world utility.
\section{Related Work}
\begin{table}[!ht]
\footnotesize
\setlength{\tabcolsep}{1.5pt}
\renewcommand{\arraystretch}{1.5}
\caption{Overview of representative PPG-based \bglEst{} approaches. Only the performance of the best model from each work is mentioned here. Methods were considered as having a random train--test split unless explicitly clarified in the work. Reported metrics are not directly comparable due to differences in experiment protocols, subject inclusion criteria, number of subjects and ML approaches across studies. Representative models evaluated in this work are highlighted.}
\centering
\begin{tabular}{
p{2cm}                  
l                  
l                  
l                  
p{1cm}                   
l                  
l                  
l                  
l                  
l                  
l                  
l                  
l                  
p{1cm}                  
}

\textbf{Work} &
\textbf{Sensor} &
\textbf{Channel} &
\textbf{Subjects} &
\textbf{BGL Range} &
\textbf{BGL M} &
\textbf{BGL SD} &
\textbf{Method} &
\textbf{MAE} &
\textbf{R$^2$} &
\textbf{MARD} &
\textbf{CEG A+B} &
\textbf{CEG A} &
\textbf{Split} \\ 

 &
 &
 &
 & 
\textbf{(mg/dL)} &
\textbf{(mg/dL)} &
\textbf{(mg/dL)} &

 &
\textbf{(mg/dL)} &
 &
\textbf{(\%)} &
\textbf{(\%)} &
\textbf{(\%)} &
 \\

\toprule

\textbf{\citet{zeynali2025non}} &
\textbf{PPG} &
\textbf{1} &
\textbf{6388} &
\textbf{12--483} &
\textbf{120.60} &
\textbf{35.80} &
\textbf{ResNet-34} &
\textbf{18.54} &
\textbf{0.40} &
\textbf{15.53} &
\textbf{98.50\%} &
\textbf{72.60\%} &
\textbf{Random} \\

\textbf{\citet{gupta2021towards}} &
\textbf{PPG} &
\textbf{1} & 
\textbf{26} &
\textbf{84--199} &
\textbf{119} &
\textbf{29} &
\textbf{XGBR} &
\textbf{8.31} &
\textbf{-} &
\textbf{-} &
\textbf{100\%} &
\textbf{96.15\%} &
\textbf{LOPO +\newline Random \newline Personali\newline -zation} \\

\textbf{\citet{alghlayini2023photoplethysmography}} &
\textbf{PPG} &
\textbf{1} & 
\textbf{52} &
\textbf{68--211} &
\textbf{-} &
\textbf{-} &
\textbf{CNN}&
\textbf{17.10} &
\textbf{-} &
\textbf{-} &
\textbf{100\%} &
\textbf{89.28\%} &
\textbf{Random} \\

\textbf{\citet{islam2021blood}} &
\textbf{Video/PPG} &
\textbf{1} &
\textbf{52} & 
\textbf{68--211} &
\textbf{-} &
\textbf{-} &
\textbf{PLS} &
\textbf{-} &
\textbf{-} &
\textbf{-} &
\textbf{-} &
\textbf{-} &
\textbf{Random,\newline LOPO} \\

\textbf{\citet{prabha2022intelligent}} &
\textbf{PPG} &
\textbf{1} &
\textbf{217} &
\textbf{58.60--390.70} &
\textbf{114.06} &
\textbf{50} &
\textbf{XGBR} &
\textbf{1.76} &
\textbf{0.99} &
\textbf{-} &
\textbf{100\%} &
\textbf{98.97\%} &
\textbf{LSPO} \\

\citet{chu202190} &
PPG,HbA1c &
1 &
2538 &
- &
- &
- &
CNN &
18.90 &
0.42 &
- &
98.5\% &
76.90\% &
Random\\

\citet{adiguzel2024blood} &
PPG &
1&
217 & 
- &
- &
- &
CatBoost &
25.16&
0.71 &
- &
- &
- &
Random \\

\citet{alghlayini2024bayesian} &
PPG &
1 & 
52 &
68 - 211 &
- &
- &
CNN &
16.91 &
- &
- &
100\% &
92.85\% &
Random \\

\citet{golap2021hemoglobin} &
Video/PPG &
1 &
111 &
48.11$^{\dagger}$--380.36$^{\dagger}$ &
116.58$^{\dagger}$ &
49.73$^{\dagger}$ &
MGGP$^{*}$ & 
0.32 &
0.88 &
- &
- &
- &
Random \\

\citet{habbu2019estimation} &
PPG &
1 &
611 &
70--450 &
- &
- &
NN &
- &
0.91 &
- &
100\% &
83\% &
Random \\

\citet{salamea2019database} &
PPG &
1 &
217 &
58.60--390.70 &
114.05 &
50.13 &
RFR &
8.59 &
- &
- &
- &
- &
Random \\

\citet{khan2026non} &
PPG &
1 &
80 &
65--160 &
- &
- &
RFR &
4.80 &
0.92 &
- &
- &
- &
Random \\ 

Hammour  &
PPG &
1 &
4 &
62--345.70 &
- &
- &
ERT$^{*}$&
- &
0.37 &
14.5 &
100\% &
82.05\% &
LSSO$^{**}$ \\

\& Mandic \cite{hammour2023ear} &&&&&&&&&&&&&\\

\citet{hossain2019estimation} &
PPG &
1 &
30 &
- &
- &
- &
CNN &
- &
- &
- &
- &
- &
LOSO$^{**}$\\

\citet{gupta2020vivo} &
PPG &
2 &
- &
- &
90/105$^{\S}$ &
- &
RFR &
- &
0.91 &
- &
- &
- &
Random \\

\citet{jia2025non} &
PPG &
2 &
20 &
- &
- &
- &
EBTA$^{*}$ & 
18.40$^{\dagger}$ &
0.79 &
- &
100\% &
80.71\% &
Random \\

\citet{jian2025using} &
PPG &
2 &
1 &
- &
- &
- &
XGBR &
- &
- &
5.15 &
100\% &
100\% &
Random \\

\citet{lee2023noninvasive} &
PPG &
2 &
18 &
- &
- &
- &
LR &
- &
- &
- &
100\% &
100\% &
LSPO \\

\citet{wang2025design} &
PPG &
2 &
1 &
63.06$^{\dagger}$--162.14$^{\dagger}$  &
- &
- &
MLP &
13.50$^{\dagger}$ &
0.71 &
- &
99.33\% &
96\% &
Random \\

\citet{castro2023three} &
PPG &
3 &
187 &
79--134 &
102 &
- &
SVM &
- &
- &
6.99 &
100\% &
95.38\% &
LSPO \\

\end{tabular}
\vspace{1.5pt}
\footnotesize
$^{*}$ EBTA - ensemble bagged trees algorithm,  MGGP - multigene genetic programming, ERT - Ensemble Regression Trees.
LOPO - Leave One Participant Out, LSSO - Leave Some Sessions Out, LSPO - Leave Some Participants Out.
$^{\dagger}$ Values converted from mmol/L to mg/dL. $^{\S}$ Pre-prandial / post-prandial values. $^{**}$ Within-subject
\label{tab:rw}
\end{table}

Non-invasive PPG-based \bglEst{} has attracted substantial research interest, driven by the promise of continuous, painless glucose monitoring. Prior work has explored sensing modalities ranging from single-channel fingertip pulse oximeters \cite{monte2011non}, multi-channel PPG \cite{castro2023three, jia2025non, lee2023noninvasive}, wrist-worn wearables \cite{prabha2022intelligent, salamea2019database}, to PPG extracted from smartphone cameras \cite{golap2021hemoglobin, islam2021blood}. In parallel, several different ML pipelines have been evaluated, including extracting time- and frequency-domain features from the PPG signal \cite{gupta2021towards, prabha2022intelligent, hammour2023ear} or directly using raw PPG signals \cite{zeynali2025non, alghlayini2023photoplethysmography, chu202190}. To improve data quality, prior work applies extensive preprocessing, ranging from filtering \cite{prabha2022intelligent,gupta2021towards, jian2025using} and motion-artifact reduction \cite{zhang2020noninvasive} to peak alignment \cite{zeynali2025non, hossain2019estimation}.

\autoref{tab:rw} summarizes representative \bglEst{} approaches. A key limitation of existing work is that most studies rely on self-collected datasets. As a result, there is wide variation in study protocols, cohort sizes, glucose ranges, and participant selection criteria, making it difficult to directly compare methods and understand which design choices actually drive performance. This highlights the need for a uniform evaluation system that allows different ML models to be implemented and evaluated across multiple publicly available datasets under standardized conditions.

Further, most studies report both ML metrics and clinical validity using Clarke Error Grid (CEG) analysis. Many papers report almost \(100\%\) of predictions in clinically acceptable zones, which, at first glance, appears highly promising for non-invasive \bglEst{}. However, closer inspection of the ML implementations reveals that train–test splits are often performed at the segment level, so windows from the same recording or even overlapping windows can appear in both sets. Next, using the same dataset for training the model, tuning hyperparameters, and reporting performance without a dedicated validation and independent test set can lead to optimistic performance estimates, a concern acknowledged in early PPG-based \bglEst{} work \cite{monte2011non}. Such data leakage inflates reported performance and makes it difficult to assess a method's true effectiveness and generalization. This motivates evaluating models under multiple splitting strategies: random segment-level splits, splits that provide explicit participant context to the model, and strict leave-some-participants-out (LSPO) protocols. Across these regimes, we systematically analyze how model reliability and conventional ML metrics relate to CEG-based clinical validity.

\section{Background}
This section introduces the key concepts and evaluation metrics that form the basis of this work.

\subsection{PPG-based \bglEst{}}
Non-invasive blood glucose monitoring with PPG relies on two principles: glucose-dependent changes in light absorption/reflection, and hemodynamic shifts that alter the pulsatile PPG waveform with changing glucose levels \cite{jiang2025ppg}. These subtle and complex patterns captured by the PPG signal can be detected by ML algorithms but often escape conventional analysis methods \cite{zeynali2025non}. 
However, PPG signal features can vary significantly due to individual differences in skin tone, blood volume, and tissue composition~\cite{jiang2025ppg}. This is the key reason why models trained and tested on the same participants may not generalize to unseen individuals, and a central motivation for the evaluation protocol presented in this work.

\subsection{Mean Absolute Relative Difference (MARD)}
MARD summarizes the average relative error between a glucose measurement system and a reference value. It is a common benchmark for comparing CGM and blood glucose monitoring systems because it provides a single measure across the full measurement range \cite{reiterer2017significance}. Unlike Mean Absolute Error (MAE), which is in absolute units, MARD reports error in relative terms, making it more comparable across glucose levels and devices. 


\subsection{Clarke Error Grid (CEG)}

The Clarke Error Grid assesses the clinical accuracy of glucose measurements by comparing device readings against reference values \cite{clarke2005original}. It sorts readings into five risk zones: A (accurate), B (acceptable), C (risk of unnecessary treatment), D (dangerous, missed hypo/hyperglycemia), and E (wrong treatment). Zones A and B are clinically acceptable, while C, D, and E denote progressively worse risk.

\section{Data}
Our pipeline incorporates two publicly available physiological datasets: \Vitaldb{} \cite{lee2022vitaldb} and \must{} \cite{kermani2023dataset}.

The \Vitaldb{} dataset contains multimodal physiological data collected at Seoul National University Hospital in Seoul, Korea during routine or emergency non-cardiac surgeries. Among various recorded signals, PPG was captured using a TramRac-4A monitor with a sampling rate of 500 Hz, while BGL were measured in the clinical laboratory. Both PPG and BGL were available for 4,872 cases with 4,626 unique participants. Patients ranged in age from 0.4 to 94 years (M = 58.32, SD = 14.78) and included 2453 men and 2173 women. The preoperative glucose levels ranged from 44 to 525~mg/dL (M = 117.20, SD = 44.27).

The \must{} dataset \cite{kermani2023dataset} was designed to facilitate research on estimating blood glucose levels from PPG signals. It contains PPG recordings sampled at 2175 Hz from 23 individuals, totaling 67 sessions. Participants were 22–61 years old (M = 33, SD = 11.17) with an average BMI of 23.26. Blood glucose values ranged from 88 mg/dL to 183 mg/dL (M = 115.01, SD = 18.67). The dataset includes 15 male and 8 female participants.
\section{Pipeline}
We designed a reproducible evaluation pipeline \cite{pipeline} to enable a fair comparison of representative PPG-based \bglEst{} models. Predictive performance was evaluated using MAE and R$^2$, and to assess clinical validity, we additionally report MARD and CEG analysis. The pipeline standardizes dataset preparation, model training, and evaluation across multiple reproduced methods. 

As mentioned before, most published works used random train--test splits, leading to data leakage. To understand the actual model performance, clinical validity, and practical relevance, we evaluated the models under three data-split protocols on VitalDB and additionally under an OOD setting, described below. Hyperparameter optimizations were performed on the respective 15\% validation set held out from the training data, while the test set was used only for evaluation. A validation set was used even when the original implementation did not have one. Each glucose value is aligned to a lab measurement timestamp, a 16-minute event-centered waveform segment, and downstream sub-windows inherit that event label. For the three \Vitaldb{} splits, we performed five-fold cross-validation and report the mean and standard deviation across folds.

\paragraph{Random Train--Test Split} The samples were randomly split at an 80-20 ratio at the PPG window level, so the windows of the same recording or even overlapping windows could appear in the train, validation, and test sets.

\paragraph{Participant-aware Train--Test Split} Prior work has demonstrated that personalization is critical for PPG-based blood pressure estimation, as subject-specific models consistently outperform general predictive approaches \cite{slapnivcar2019blood}. To evaluate model performance under this participant-aware setting, we designed a train--test split targeting participants with multiple recordings. Of the participants who had more than one session, exactly two recordings per participant were retained to ensure a fair and balanced comparison. One was assigned to the train set and one to the test set. Participants with only a single recording were allocated entirely to the train set. We conducted five-fold cross-validation with non-overlapping participant assignments across folds. In each fold, the participants selected for evaluation contribute one session to the training set and one to the test set, while the remaining participants with multiple recordings are included solely in the training set. This protocol directly measures how well the model generalizes when it has already observed data from the target individual. Since only 76 participants had multiple recordings in the dataset, this analysis only uses them for evaluation.

\paragraph{LSPO Train--Test Split} Train and test sets were split to have samples from non-overlapping participants, with the test set having one-fifth of the total participants. With this analysis, we evaluate how the model reacts to unseen participants' data.

\paragraph{Out-of-Distribution (OOD) Generalization} \must{} dataset due to its small size could not be used for training the deep learning models but was instead used to assess OOD generalization \cite{zeynali2025non}. For this, we trained the model with all the \Vitaldb{} data and tested generalization on the \must{} dataset.

\subsection{Evaluated Models}
We chose five representative works from the literature that used PPG signal for \bglEst{}. The models were chosen based on the clarity of their methodological descriptions and their compatibility with the single-channel PPG data available in the datasets used in this work. Since most works presented more than one ML architecture, we chose the model that performed best in each study. All the preprocessing steps were retained based on the descriptions from their original studies to enable a fair comparison. Since none of the works published their code, we built the models based on the details presented in their work. Along with the representative models, baseline results were evaluated by using the mean of the training set as the prediction.

\paragraph{Raw PPG and Resnet-34 Model \cite{zeynali2025non}}
This work proposed deep learning models for non-invasive \bglEst{} using raw PPG signals from the \Vitaldb{} dataset, with additional testing on the \must{} dataset. Their preprocessing pipeline included downsampling, temporal alignment of PPG and glucose measurements, filling missing values, bandpass filtering, and segmentation around characteristic waveform peaks. For the model, they presented a random 70-15-15 split. We selected ResNet-34 as the representative implementation for our pipeline since it was their best-performing model.  Clinical evaluation in the original study included CEG and MARD, which we also report in our experiments.

\paragraph{Raw PPG and 1D-CNN Model \cite{alghlayini2023photoplethysmography}}
This work proposed a lightweight 1D-CNN model evaluated on 20-second fingertip PPG signals acquired with a pulse oximeter sensor. A low-pass Butterworth filter was used to remove noise. For the model, a random 85-15 split was used with no separate validation set. MAE was used as the training metric, and clinical validity was assessed using CEG analysis.

\paragraph{MFCC Feature Extraction and XGBoost Model \cite{prabha2022intelligent}}
This work proposed an XGBoost model evaluated on wristband PPG signals. Preprocessing included segmentation into 5-second windows, removal of noisy segments, bandpass filtering, MFCC-based feature extraction, and z-score normalization. The extracted MFCC features were combined with demographic variables, and feature selection was performed using XGBoost importance scores. The original study evaluated the optimized model using both LSPO split and ten-fold cross-validation. No separate validation set was mentioned. Clinical validity was assessed using CEG analysis and MARD. 

\paragraph{Feature Extraction and Partial Least Squares Model \cite{islam2021blood}}
This work proposed a smartphone-based \bglEst{} method evaluated on 60-second fingertip videos. The PPG signal was obtained from the red channel of the recorded video frames and was preprocessed using asymmetric least squares for baseline correction, Gaussian low-pass filtering, and normalization. Then, temporal and derivative-based features were extracted from the waveform. For model building and independent testing, the data was randomly split in a 75-25 ratio without prior stratification. Five- and ten-fold cross-validation were performed, evaluating four regression techniques. However, no separate validation set was mentioned. Partial Least Squares (PLS) model achieved the best performance. 

\paragraph{Feature Extraction and XGBoost Model \cite{gupta2021towards}}
This work proposed a PPG-based \bglEst{} approach using a custom sensing system that recorded finger PPG signals over four-minute intervals. Their preprocessing pipeline removed baseline drift using a fitting-based sliding window method, low-pass Butterworth filtering, and segmented the signals into 3-second windows. Signal quality was screened using heart rate and breathing rate criteria, and handcrafted spectral, statistical, and physiological features were then extracted from each segment. The original study evaluated Random Forest and XGBoost regressors and validated performance using a leave-one-out setting with partial personalization, where a random 1\% of the target subject's data was included in training. No separate validation set was mentioned. Although the paper briefly mentions multiple wavelengths, the feature descriptions and reported analyses do not clearly distinguish channel-specific contributions except for SpO$_2$; therefore, for reproducibility, we treated this method as a single-channel approach in our implementation.

\begin{table}[!ht]
\setlength{\tabcolsep}{2.25pt}
\renewcommand{\arraystretch}{1.15}
\footnotesize
\caption{Results of the representative models across all evaluations. }
\centering
\begin{tabular}{llll|llllll}
\multirow{2}{*}{\textbf{Model}} & \multirow{2}{*}{\textbf{Data Split}} & \multirow{2}{*}{\textbf{MAE (mg/dL)}} & \multirow{2}{*}{\textbf{R$^2$}} & \multirow{2}{*}{\textbf{MARD (\%)}} & \multicolumn{5}{c}{\textbf{CEG (\%)}} \\
& & & & & \textbf{Zone A} & \textbf{Zone B} & \textbf{Zone C} & \textbf{Zone D} & \textbf{Zone E} \\
\toprule
Global Baseline & Random & 26.40 $\pm$ 0.50 & -0.00 $\pm$ 0.00 & 22.57 $\pm$ 0.61 & 52.03 $\pm$ 1.09 & 45.38 $\pm$ 0.79 & 0.00 $\pm$ 0.00 & 2.59 $\pm$ 0.33 & 0.00 $\pm$ 0.00 \\
Global Baseline & Participant-aware & 30.61 $\pm$ 6.98 & -0.10 $\pm$ 0.16 & 22.18 $\pm$ 3.54 & 50.61 $\pm$ 4.82 & 46.76 $\pm$ 3.00 & 0.00 $\pm$ 0.00 & 2.63 $\pm$ 2.55 & 0.00 $\pm$ 0.00 \\
Global Baseline & LSPO & 26.39 $\pm$ 1.08 & -0.00 $\pm$ 0.00 & 22.52 $\pm$ 0.96 & 52.27 $\pm$ 1.50 & 45.14 $\pm$ 1.36 & 0.00 $\pm$ 0.00 & 2.60 $\pm$ 0.42 & 0.00 $\pm$ 0.00 \\
Global Baseline & OOD & 16.00 & -0.08 & 14.13 & 77.61 & 22.39 & 0.00 & 0.00 & 0.00 \\
\midrule
\citet{zeynali2025non} & Random & 13.97 $\pm$ 0.20 & 0.60 $\pm$ 0.00 & 12.12 $\pm$ 0.21 & 82.17 $\pm$ 0.54 & 16.62 $\pm$ 0.50 & 0.07 $\pm$ 0.01 & 1.13 $\pm$ 0.05 & 0.00 $\pm$ 0.00 \\
\citet{zeynali2025non} & Participant-aware & 36.18 $\pm$ 4.51 & -0.08 $\pm$ 0.22 & 30.97 $\pm$ 7.53 & 39.42 $\pm$ 8.37 & 52.42 $\pm$ 10.71 & 0.01 $\pm$ 0.01 & 8.13 $\pm$ 11.56 & 0.03 $\pm$ 0.06 \\
\citet{zeynali2025non} & LSPO & 25.93 $\pm$ 2.16 & -0.04 $\pm$ 0.03 & 22.55 $\pm$ 1.02 & 55.62 $\pm$ 3.50 & 41.43 $\pm$ 3.09 & 0.04 $\pm$ 0.03 & 2.90 $\pm$ 0.73 & 0.01 $\pm$ 0.00 \\
\citet{zeynali2025non} & OOD & 16.33 & -0.28 & 14.99 & 72.17 & 27.83 & 0.00 & 0.00 & 0.00 \\
\midrule
\citet{alghlayini2023photoplethysmography} & Random & 26.02 $\pm$ 0.47 & 0.00 $\pm$ 0.01 & 21.94 $\pm$ 0.62 & 53.98 $\pm$ 1.24 & 43.44 $\pm$ 1.13 & 0.00 $\pm$ 0.00 & 2.59 $\pm$ 0.12 & 0.00 $\pm$ 0.00 \\
\citet{alghlayini2023photoplethysmography} & Participant-aware & 30.57 $\pm$ 7.38 & -0.12 $\pm$ 0.17 & 21.94 $\pm$ 4.06 & 50.64 $\pm$ 11.06 & 46.72 $\pm$ 9.13 & 0.00 $\pm$ 0.00 & 2.63 $\pm$ 2.55 & 0.00 $\pm$ 0.00 \\
\citet{alghlayini2023photoplethysmography} & LSPO & 26.05 $\pm$ 1.23 & -0.01 $\pm$ 0.01 & 21.80 $\pm$ 1.31 & 54.19 $\pm$ 2.47 & 43.22 $\pm$ 2.05 & 0.00 $\pm$ 0.00 & 2.60 $\pm$ 0.42 & 0.00 $\pm$ 0.00 \\
\citet{alghlayini2023photoplethysmography} & OOD & 18.21 & -0.34 & 16.56 & 62.69 & 37.31 & 0.00 & 0.00 & 0.00 \\
\midrule
\citet{prabha2022intelligent} & Random & 23.98 $\pm$ 0.80 & 0.16 $\pm$ 0.06 & 20.40 $\pm$ 0.68 & 58.17 $\pm$ 1.73 & 39.28 $\pm$ 1.71 & 0.00 $\pm$ 0.00 & 2.55 $\pm$ 0.05 & 0.00 $\pm$ 0.00 \\
\citet{prabha2022intelligent} & Participant-aware & 29.66 $\pm$ 5.06 & -0.03 $\pm$ 0.08 & 21.91 $\pm$ 3.29 & 55.25 $\pm$ 4.75 & 42.12 $\pm$ 3.40 & 0.00 $\pm$ 0.00 & 2.63 $\pm$ 2.55 & 0.00 $\pm$ 0.00 \\
\citet{prabha2022intelligent} & LSPO & 25.42 $\pm$ 1.01 & 0.05 $\pm$ 0.01 & 21.56 $\pm$ 0.85 & 55.10 $\pm$ 2.10 & 42.30 $\pm$ 2.01 & 0.00 $\pm$ 0.00 & 2.60 $\pm$ 0.42 & 0.00 $\pm$ 0.00 \\
\citet{prabha2022intelligent} & OOD & 14.23 & -0.06 & 11.69 & 83.58 & 16.42 & 0.00 & 0.00 & 0.00 \\
\midrule
\citet{islam2021blood} & Random & 26.18 $\pm$ 0.19 & 0.01 $\pm$ 0.00 & 22.40 $\pm$ 0.09 & 52.98 $\pm$ 0.26 & 44.42 $\pm$ 0.20 & 0.00 $\pm$ 0.00 & 2.59 $\pm$ 0.15 & 0.00 $\pm$ 0.00 \\
\citet{islam2021blood} & Participant-aware & 29.56 $\pm$ 6.12 & -0.05 $\pm$ 0.08 & 21.67 $\pm$ 3.25 & 52.64 $\pm$ 7.39 & 44.72 $\pm$ 6.14 & 0.00 $\pm$ 0.00 & 2.63 $\pm$ 2.55 & 0.00 $\pm$ 0.00 \\
\citet{islam2021blood} & LSPO & 26.14 $\pm$ 1.17 & 0.01 $\pm$ 0.01 & 22.28 $\pm$ 0.97 & 53.19 $\pm$ 1.83 & 44.21 $\pm$ 1.65 & 0.00 $\pm$ 0.00 & 2.60 $\pm$ 0.42 & 0.00 $\pm$ 0.00 \\
\citet{islam2021blood} & OOD & 13.37 & -0.10 & 10.67 & 86.57 & 13.43 & 0.00 & 0.00 & 0.00 \\
\midrule
\citet{gupta2021towards} & Random & 22.40 $\pm$ 0.03 & 0.27 $\pm$ 0.00 & 19.20 $\pm$ 0.04 & 60.99 $\pm$ 0.13 & 36.61 $\pm$ 0.13 & 0.00 $\pm$ 0.00 & 2.40 $\pm$ 0.02 & 0.00 $\pm$ 0.00 \\
\citet{gupta2021towards} & Participant-aware & 29.18 $\pm$ 6.24 & -0.00 $\pm$ 0.08 & 21.68 $\pm$ 4.07 & 56.89 $\pm$ 7.24 & 40.49 $\pm$ 6.17 & 0.00 $\pm$ 0.00 & 2.62 $\pm$ 2.53 & 0.00 $\pm$ 0.00 \\
\citet{gupta2021towards} & LSPO & 25.77 $\pm$ 1.09 & 0.04 $\pm$ 0.01 & 21.96 $\pm$ 0.92 & 53.88 $\pm$ 1.65 & 43.54 $\pm$ 1.45 & 0.00 $\pm$ 0.00 & 2.58 $\pm$ 0.44 & 0.00 $\pm$ 0.00 \\
\citet{gupta2021towards} & OOD & 16.69 & -0.15 & 14.46 & 75.60 & 24.40 & 0.00 & 0.00 & 0.00 \\
\end{tabular}
\label{tab:results}
\end{table}

\begin{figure*}[!ht]
\centering
\captionbox{%
    CEG analysis for \citet{prabha2022intelligent}, pooled across all five folds; color indicates prediction density. Predictions concentrate around the training mean under all three splits, consistent with the near-zero R$^2$ in \autoref{tab:results}.%
    \label{fig:ceg_prabha}
}{%
    \includegraphics[width=0.48\linewidth]{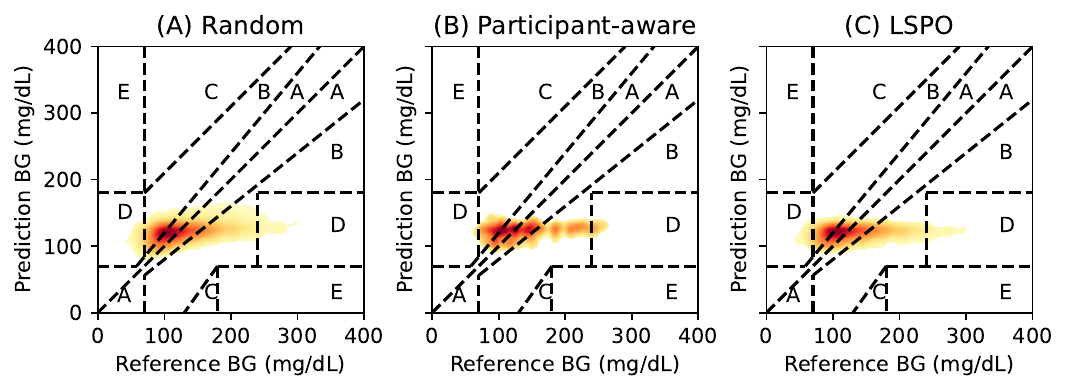}
}%
\hfill
\captionbox{%
    CEG analysis for \citet{zeynali2025non}, pooled across all five folds; color indicates prediction density. The prediction spread narrows sharply from (A) to (B) and (C), with most values concentrated near the mean glucose range.%
    \label{fig:ceg_zeynali}
}{%
    \includegraphics[width=0.48\linewidth]{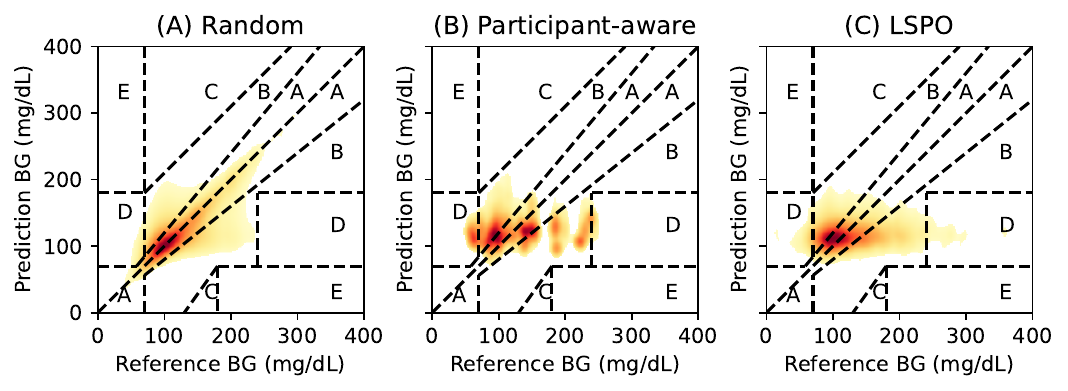}
}

\Description{Plot has two subfigures, right and left. The left side has three Clarke Error Grid plots for Prabha et al. under random, participant-aware, and LSPO splits. Across all three splits, predictions form a dense horizontal band concentrated around 80--120~mg/dL on the predicted axis, with no diagonal trend tracking the reference glucose values. The clustering pattern remains consistent across all three evaluation settings, showing no meaningful difference between splits. The majority of predictions fall within Zone~A and Zone~B, with negligible contributions from Zones~C, D, and~E. The right side has three Clarke Error Grid plots for Zeynali et al. under random, participant-aware, and LSPO. The random split shows a diagonal spread of predictions tracking reference glucose values. The participant-aware and LSPO splits show predictions collapsed into a narrow horizontal band around 80--120~mg/dL, with no tracking of reference values. Most predictions fall within Zone~A and Zone~B across all splits.}
\end{figure*}

\section{Results}

In this section, we present the results of the five representative models across three \Vitaldb{} splits and one OOD evaluation. Differences in data preprocessing and cleaning steps across methods make direct cross-model comparison difficult; however, the performance of each model across data splits can be meaningfully assessed. \autoref{tab:results} shows the results of representative models.

\paragraph{ML Performance}
Across all five models, MAE increases, and R$^2$ drops when moving from the random window-level split to the participant-aware split, with every model yielding near-zero or negative R$^2$ under the latter. For every model, performance under the strict LSPO split is better than under the participant-aware split. Notably, only \citet{prabha2022intelligent}, \citet{gupta2021towards}, and \citet{islam2021blood} maintain a positive R$^2$ under the LSPO split. \citet{zeynali2025non} achieves the highest R$^2$ of $0.60\pm0.00$ under the random split, yet yields R$^2 = -0.08\pm0.22$ and $-0.04\pm0.03$ under the participant-aware and LSPO splits, respectively, representing the sharpest performance degradation across all evaluated models. Baseline MAE is similar to the rest of the model performance except for the random split, with R$^2$ always around zero or lower. The participant-aware split shows the largest fold-to-fold variance because it is restricted to the 76 participants. Its point estimates are therefore noisier and not directly comparable to the LSPO split, which uses the full cohort. Even so, the central pattern holds: prior exposure to the target individual yields no reliable improvement over the unseen setting.

\paragraph{CEG and MARD Analysis}
Despite poor regression performance, the majority of predictions across all models and all three splits fall within Zone A and Zone B of the CEG. MARD values remain largely stable across the three splits for most models. The \citet{zeynali2025non} implementation is an exception, with MARD rising from $12.12\pm0.21\%$ under the random split to $30.97\pm7.53\%$ under the participant-aware split — the largest shift observed across all models. Importantly, the CEG and MARD values of the baseline align with the results of the other models. The divergence between stable MARD and collapsing R$^2$ across splits implies that models maintain a consistent magnitude of error by learning where most train sample glucose ranges lie but lose all ability to capture glucose variance across individuals. As visible in \autoref{fig:ceg_prabha} and \autoref{fig:ceg_zeynali}, predicted glucose values are concentrated within a narrow range irrespective of the reference glucose value, forming a horizontal band rather than the diagonal pattern expected of an accurate predictor. 

\paragraph{OOD Generalization} The \must{} dataset has a very small glucose distribution $115.01\pm18.67$ mg/dL, close to VitalDB's mean ($117.20\pm44.27$ mg/dL). Therefore, when the model predicted values around the mean, it achieved a lower MAE, and all results fell within the CEG clinically acceptable zones, while the R$^2$ remained below zero for all models.





\section{Discussion}
Our results demonstrate that reported performance in PPG-based BGL estimation is strongly driven by evaluation methodology rather than true model generalization. All five reproduced methods degraded to mean-baseline levels under stricter data splits, yet the clinical validity metric concealed this failure entirely. This underscores the need to critically examine both how models are evaluated and how performance is reported.

\begin{table}[!ht]
\footnotesize
\setlength{\tabcolsep}{2pt}
\renewcommand{\arraystretch}{2}
\centering
\caption{Evaluation metrics used in this work, with what each captures and its blind spot.}
\label{tab:metrics_reference}
\begin{tabular}{l p{3cm} p{5cm} p{5cm}}

\textbf{Metric} & \textbf{Definition} & \textbf{Captures} & \textbf{Blind spot} \\
\toprule
MAE &
Average absolute error &
Easy to interpret since it is in the target's original units, robust to outliers &
Error directionality, conflates systematic and random error, scale-dependent \\
$R^2$ &
Variance explained relative to a mean baseline &
Detects if predictions track real variability, not just the mean &
Depends on the spread of the data \cite{shalizi2013advanced}: R$_2$ can fall towards zero if the predicted value range is narrow \\
MARD &
Absolute error as percentage of true value &
Standard for cross-study glucose comparisons &
Similar values can have different clinical consequences; since the error is normalized by the true value, MARD tends to be higher at low glycemic ranges \cite{heinemann2020benefits}.\\
CEG &
Clinical risk zone of each prediction &
Directly reflects treatment/safety risk &
Can hide poor regression performance behind ``safe'' zones \\

\end{tabular}
\end{table}

\paragraph{Data leakage and dependence between samples.}
A key finding is that model performance strongly depends on the evaluation protocol: under random train--test splits, all models showed lower MAE and higher R$^2$ than the mean baseline, but these gains largely disappeared under participant-aware and LSPO splits. This discrepancy is consistent with sample dependence or data leakage, since samples from the same recording or participant often share context- and subject-specific characteristics, artificially easing prediction when correlated samples appear in both training and test sets \cite{karbalaie2026participant, kapoor2023leakage}. Our findings show that random-split evaluation overestimates these models' real-world deployment utility.

\paragraph{Dedicated validation set} Having a separate validation set is essential to avoid further data leakage due to hyperparameter optimization and evaluation on the same data \cite{kapoor2023leakage}. This could explain the performance differences across all representative models, except \citet{zeynali2025non}, even with a random train--test split.

\paragraph{Clinical validity and regression performance.} Another key observation is the discrepancy between clinical error-grid metrics and standard regression metrics: over 90\% of predictions fell within Zones A and B across all settings, even at the mean baseline, yet MAE and MARD remained high with near-zero or negative R$^2$ under participant-aware and LSPO splits. This shows that zone-based metrics alone are insufficient and should be interpreted alongside performance measures and baseline comparisons. \autoref{tab:metrics_reference} summarizes the four metrics discussed in this work and their purpose and blind spot. A more balanced dataset across the glucose range could help models learn more effectively and better reflect this in clinical metrics; future studies should explore data augmentation or upweighting samples farther from the mean glucose level to improve performance.

\paragraph{Scope for patterns in handcrafted features} Among the representative models, we could observe that both the deep learning models that consider raw time-domain PPG signals failed to have a good generalization to unseen data. In contrast, the models that used handcrafted feature extraction methods achieved better R$^2$ on LSPO. Further work should investigate the features contributing to this performance.

\paragraph{PPG-based \bglEst{}.}
Taken together, the results suggest that \bglEst{} from PPG alone remains a challenging task, especially since the participant-aware split, where the model has already seen the target individual, performs no better than the fully unseen LSPO split. The reproduced models did not demonstrate robust generalization, which limits their immediate practical applicability. This may indicate that improved preprocessing, stronger control of confounding factors, and the inclusion of additional sensing modalities are necessary to obtain more reliable and clinically useful estimates.

\subsection{Limitations}

\paragraph{Reproduction from descriptions} Since our evaluation pipeline was implemented solely based on each work's methodology descriptions, variation in results can occur due to differences in model training and tuning across datasets and gaps in the descriptions. 

\paragraph{Channel ambiguity in \citet{gupta2021towards}} The mapping of PPG channels to features was not mentioned in \citet{gupta2021towards} except for SpO$_2$. We used the SpO$_2$ measurements from Solar8000 directly for \Vitaldb{} evaluations. In the OOD evaluation, the SpO$_2$ feature was dropped since \must{} has a single-channel PPG and does not have separate SpO$_2$ measurements.

\paragraph{Perioperative data} \Vitaldb{} dataset was collected in perioperative conditions. Although this dataset has been used in two earlier works \cite{zeynali2025non, young2025low}, the subjects are measured during surgery and therefore under anesthesia, which could impact the data.

\subsection{Future Directions}
The results of this work also point to concrete opportunities to extend the evaluation pipeline and advance the field toward more reliable and generalizable PPG-based \bglEst{}.

\paragraph{Extension of the pipeline}  This study is limited to a set of representative models and the evaluation protocols applied to the available datasets. More published datasets need to be integrated into the pipeline and used to advance PPG-based \bglEst{} research. Further, we have also implemented the model size evaluation in the pipeline to understand the utility of the models in the edge device context. In this work, we didn't discuss these results since none of the representative models performed well. Further, it has already been established that PPG signal quality depends on the measurement location \cite{hartmann2019quantitative}. This implementation can be used to evaluate how blood glucose measurements vary with PPG collected across different body sites.

\paragraph{Identifying the root cause} Future work should investigate more robust modeling strategies, frequency domain-based deep learning approaches, data augmentation strategies, and multimodal approaches that combine PPG with other physiological signals. Such directions could help clarify whether the current limitations arise primarily from model design, dataset characteristics, or the intrinsic difficulty of non-invasive glucose estimation from PPG.

\section{Conclusion}
This work presents the first reproducible and extendable pipeline for PPG-based \bglEst{} and compares the performance of five representative models on two published datasets. Three different train--test data splits were evaluated. Results show that current research overestimates the generalization of ML models due to multiple levels of data leakage. Additionally, we show that ML approaches for \bglEst{} studies should focus more on implementing a robust ML pipeline that follows best practices before evaluating clinical metrics, to develop applications with practical utility. Future research should validate the viability of PPG-based \bglEst{} applications and also develop generalizable models.

\begin{acks}
This work is partially funded by the KIT Center of Health Technologies, by the program Core-Informatics of the Helmholtz Association (HGF: KIKIT/COIN), and the HEiKA funded project PACo, and supported by the Helmholtz Association Initiative and Networking Fund on the HAICORE@KIT partition.
\end{acks}

\bibliographystyle{ACM-Reference-Format}
\bibliography{references}

\end{document}